\documentclass{WileyMSP-template}
\usepackage{bm}
\usepackage{geometry}
\usepackage{fancyhdr}
\usepackage{xcolor}
\usepackage{amsmath}
\usepackage{amssymb}

\justifying

\begin{document}

\pagestyle{fancy}
\rhead{\includegraphics[width=2.5cm]{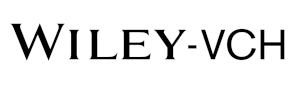}}

\title{Tomonaga-Luttinger Liquid parameters in Multi-wall Nanotubes}

\maketitle

\author{Naira Grigoryan,}
\author{Piotr Chudzinski}

\begin{affiliations}
Naira Grigoryan\\
Institute of Fundamental Technological Research, Polish Academy of Sciences, Adolfa Pawińskiego 5b, 02-106 Warsaw, Poland\\
Email Address: nairagr@ippt.pan.pl

Piotr Chudzinski\\
Institute of Fundamental Technological Research, Polish Academy of Sciences, Adolfa Pawińskiego 5b, 02-106 Warsaw, Poland and\\
School of Mathematics and Physics, Queen's University Belfast, University Road, Belfast, NI BT7 1NN, United Kingdom\\
Email Address: pchudzin@ippt.pan.pl 

\end{affiliations}

\keywords{Tomonaga-Luttinger liquid,   single-wall nanotubes, multi-wall nanotubes,  Coulomb interaction in 1D.}


\date{\today}

\begin{abstract}
Tomonaga-Luttinger liquid (TLL) theory is a canonical formalism used to describe one-dimensional (1D) metals, where the low energy physics is determined by collective bosonic excitations. In this work, we present a theoretical model to compute the parameters of Tomonaga-Luttinger liquid (TLL) in multi-wall nanotubes (MWNTs). MWNTs introduce additional complexity to the usual fermionic chains due to interactions and hybridization between their multiple coaxial shells. 
We consider a model in which conducting paths along the length of the MWNTs are randomly distributed among the shells. 
Since the valley degree of freedom remains a good quantum number, the TLL description 
in addition to spin and charge, contains also valley degree of freedom, hence four mode description applies. The values of all four TLL parameters are obtained for this model. A surprising outcome is that the compressibility of the holon mode becomes a universal quantity, while the parameters of neutral modes will depend on the details of inter-shell coupling. Finally, we propose experiments where our predictions can be tested.
 
\end{abstract}

\section{Introduction}\label{sec: Introduction}

Multi-wall carbon nanotube (MWCNT) \cite{iijima1991helical, kukovecz2013multi} is a system consisting of several interlinked single-walled carbon nanotubes (SWCNTs) \cite{mceuen2000single} with a diameter reaching more than $100$ nm. Each shell of a MWCNT is a SWCNT, which is itself a rolled sheet of graphene \cite{d2017graphene}. For this system, the theoretical description is well-known and established \cite{boumia2014timoshenko, 10.1063/1.1419236}:  one-third of them will be metallic, potentially susceptible to mini-gap opening due to curvature \cite{liu2015curvature, dumitricua2002curvature} and spin-orbit effects \cite{chudz-so-cnt, pichugin2014spin}, while the two-thirds are wide-gap semiconductors \cite{hamada1992new, dresselhaus2005semiconducting}.
Although a SWCNT is harder to produce its theoretical description is more straightforward. On the other hand, a MWCNT is easier to synthesize and produce \cite{jabeen2015review} on the industrial scale, but quite difficult to model theoretically. 

Nanotube is one of the best examples of a one-dimensional (1D) system where the motion of carriers is confined to one direction only, thus they cannot avoid each other in their motion. They form a collective state known as Tomonaga-Luttinger liquid (TLL) \cite{giamarchi2003quantum, haldane1981luttinger}, an intriguing state of matter that is nowadays still a subject of intense experimental research\cite{he2020emergence, PhysRevB.108.L201101, doi:10.1126/science.abn1719}. Instead of an effective mass of the Fermi liquid's quasi-particle, in TLL the central quantity of interest are compressibilities of bosonic modes, the TLL parameters $K_{\nu}$, and substantial effort is being made to find them\cite{shen2023effect, horvatic2020direct, 10.1093/pnasnexus/pgae363}. For MWCNT, a set of 1D systems, the situation is more unclear and obviously the theoretical description is going to be more complex but still in term of a set of $K_{\nu}$s. 

To be specific, our aim here is to derive an effective theory for MWNT not at zero temperature, but at experimentally (and technically) relevant finite temperatures\cite{lange2024metal, cavazos2023thermal}. At the lowest temperatures, one expects Coulomb blockade effects and weak-localization many-paths interference to dominate the physics\cite{EGGER2002447}. However, at higher temperatures when $T>E_C$, the electrons can start to jump freely between the large quantum dots and, based on the quasi-1D cross-over model, we know that only the strongest paths will allow for coherent states. This is the regime where collective bosonic modes (the TLL modes) can dominate the MWNT properties, a postulate supported by recent experimental findings\cite{inaba2022electron, dini2020unified}.   





When there are several coaxial tubes, as in MWCNT, their theoretical description should be a combination of SWCNT microscopic theories \cite{su2007work}. If they do not interact, it would be a simple linear combination: once hybridization is allowed, it becomes a tensor product of the constituents. Extracting properties of MWCNT become an extremely difficult task already on the single-particle level. 
From the SWCNT, we know that there are two valleys: $K$ and $K'$. Therefore, the effective description will consider the two ladders, which in the case of Luttinger liquid, will correspond to the fluctuations of not only spin and charge but also valley degree of freedom {\cite{cavazos2023thermal, egger2001luttinger}. 

One needs to construct an effective theory for these lowest band carriers. This is our first task, in Sec. \ref{sec: model}. 
In Sec. \ref{sec:KinSWCNT}, we review the TLL parameters in SWCNTs. The main results are introduced in Sec. \ref{sec:KinMWNT}: in Sec. \ref{sec:holon_mode}, we calculate the values of the parameter for the holon mode. Therefore, in  Sec. \ref{ssec:neutral-modes}, we analyze the three neutral modes and in Sec. \ref{ssec:measurK}, we discuss how these parameters can be measured experimentally. The paper concludes with Sec. \ref{Sec: Discussion and Conclusions}, where we discuss the validity of our approximation.

\section{The model}\label{sec: model}

The theoretical description of a SWCNT on a single particle level is well-known and is derived from periodic boundary conditions applied to a graphene ribbon. When a graphene ribbon is rolled into a SWCNT, periodic boundary conditions are applied along the circumference of the tube \cite{Nanot2013}. The construction of a SWCNT is determined by its chiral vector $\vec{C}_h$, which is defined in terms of the graphene lattice vectors $\vec{a}_1$ and $\vec{a}_2$:

\begin{equation}
    \vec{C}_h = n \vec{a}_1 + m \vec{a}_2,
\end{equation}

where $n$ and $m$ are the chiral indices. The chiral vector describes how the graphene sheet is rolled to form the nanotube. If $(n - m) \mod 3 = 0$, the nanotube is metallic; otherwise, the nanotube is semiconducting \cite{10.1063/1.1419236}. SWCNTs have two valleys, $K$  and $K'$. These valleys correspond to the two points in the Brillouin zone of graphene, which are preserved when the graphene sheet is rolled into a nanotube. The nanotube is a metal or a semiconductor depending on whether any of the quantized bands crosses the $K, K'$ points. 
Further improvements of the single-particle theory included spin-orbit \cite{chudz-so-cnt}, curvature \cite{liu2015curvature, dumitricua2002curvature}, and strain \cite{mishra2018coaxial, ye2023study} effects.

We are now ready to add electron-electron interactions to our model, that is to move to the TLL description in terms of collective modes. The interaction term, the two-body interactions between electrons in a nanotube, reads in general:

\begin{equation}\label{eq:int-def}
    H_{Coul}=\sum_{k,q,\mu} [c_{\mu}^{\dag}(k)c_{\mu}(k)]V_{Coul}^{\mu\mu'}(k,k')[c_{\mu'}^{\dag}(k+q)c_{\mu'}(k+q)]
\end{equation}

which we have expressed here in the fermionic second quantization language as that makes the physical content of the interaction transparent: one sees that there is a summation over all $\mu$ valley, spin, and sub-lattice states and the interaction may depend on these. In 1D, we can apply the bosonization method, whereby all fermionic operators are expressed in terms of bosonic fields:

The collective modes Hamiltonian is:

\begin{equation}\label{eq:ham-TLL-def}
    H^{1D} = \sum_{\nu}^{N} \int\frac{\mathrm{d}x}{2\pi}
    \left[(v_{\nu}K_{\nu})(\pi \Pi_{\nu})^{2}+\left(\frac{v_{\nu}}{K_{\nu}}\right)(\partial_{x} \phi_{\nu})^{2}\right]
\end{equation}

where $\nabla\phi_{\nu}(x)$ gives the local density of fluctuations, while $v_{\nu}$ and $K_{\nu}$ are respectively the velocity and the TLL parameter ($\sim$ compressibility) of a given bosonic mode $\nu$ that depend on electron-electron interactions with small momentum exchange.

The SWNT's description in terms of TLL modes will have important implications for inter-shell tunneling. It is known that the hybridization between 1D systems is renormalized by interactions\cite{CARON1988A67}:
\begin{equation}
    t_{\perp}^{eff}=(V_F \pi/a)\bigg(\frac{t_{\perp}}{V_F\pi/a}\bigg)^{\frac{1}{2-\varsigma}}
\end{equation}
where $a$  is an effective lattice constant (along the tube), and $\varsigma$ is a single particle Greens' function exponent:
\begin{equation}
    \varsigma = \sum_{\nu}^{N_\nu} \frac{\left( K_{\nu} + K_{\nu}^{-1}\right) }{2N_{\nu}}
\end{equation}
where the summation goes over all bosonic modes, and in our case of $K_{\rho+}\ll 1$, this will be a strong renormalization of $t_{\perp}$ downwards. The bare value of inter-shell hopping, for a convenient relative orientation of respective chiral angles, may reach values \cite{yu2022interlayer} up to $t_{\perp}=0.31eV$, but we see that this value is reduced by both strong interactions and large unit cells $a$, especially for the shell with largest chiral vector. In general, the $t_{\perp}^{eff}$ will be smaller than most single-particle gaps $\Delta_{sp}$ in semi-conducting SWNT, especially in the narrowest tubes in the core of MWNT. This is because $\Delta_{sp}\propto 1/R$, thus these tubes will conduct only when metallic (or when strongly distorted). On the other hand, the outer-most shells will be the most affected by interaction with the environment.

We can now move to the theoretical description of MWNT. Obviously, the theoretical description ought to be based on SWNT. With a non-hybridized, non-interacting shells, we will obtain a band structure that is a simple sum of constituting shells, but this situation is not physically relevant since some coupling must be present.  

In the presence of weak inter-shell interactions, one can consider SWNT spectra produced for strained and twisted shells \cite{mishra2018coaxial, ye2023study}. Band structure remains qualitatively the same, except for the mini-gaps that may open due to emergent back-scattering terms. Hybridization is a more complicated problem. Qualitatively, it is easy to imagine that there is a system of single particle bands that are crossing each other due to their relative shifts, varying chiralities and mini-gaps. In the presence of hybridization, one expects bands of anti-crossing to appear (see Fig. \ref{fig:MWNTmod}). The more shells, the higher the density of such anti-crossing points. Here, we shall assume that such anti-crossing is in fact a normal state of affairs in the system, that is the lowest energy band is in fact a sequence of close consecutive anti-crossings. This picture immediately explains the reduced Fermi velocity frequently observed in MWCNT, e.g. \cite{saqib2018natural}.

SWNT bands are orthogonal and they never cross. Thus, the above given reciprocal space picture of consecutive anti-crossings translates into a random sequence of inter-shell hoppings in real space. One can define the following Hamiltonian:
\begin{equation}\label{Eq: Hamiltonian_inter-shell hoppings}
    H_{\perp}= \int dx \sum_k \sum_{n,n'} \overline{\overline{t_{\perp}}}(x,n,n';v_n(x),v_{n'}(x))c_{\mu}^{\dag}(x,n,k)c_{\mu}(x,n',k) 
\end{equation}
where we assume that the momentum $k$ and other quantum numbers $\mu$ are conserved during the hopping event. It should be noted that in our model, we admit the dependence $v_n(x)$, i.e., the velocity of carriers depends on the position along the tube. In MWNT, there are several factors that can induce such dependence: i) the local strain/twist due to inter-tube interactions; ii) local atomic defect that changes tubes chirality and thus the relative position of carbon atoms of neighboring tubes; iii) inhomogeneity of electron-electron interaction due to the varying environment. The last contribution was already identified in Ref. \cite{Carr-dimSWCNT} as the one responsible for an interaction-induced dimerization in SWCNT. In MWNT, this effect is expected to be much stronger and, with several available periodicities, it will lead to Aubry-Andree type complicated renormalization of $v(x)$. From a more general point of view, one expects that the inter-tube interactions for two shells of different chiralities will be governed by an emergent Moire pattern, hence a feature that varies along the MWNT length. 

Crucially, the hopping probability $\overline{\overline{t_{\perp}}}(x,n,n';v_n,v_{n'})$ is not only a function that decays very quickly with $|n-n'|$ (which is physically reasonable, since the carrier should be able to jump only to the nearest neighboring shell), but also it is a conditional quantity. The hopping is accepted only when $v_n(x)>v_{n'}(x)$, that is when the mobility of a carrier in the new-shell is larger than in the past-shell. In this way, the model is quite restrictive in the sense that it is not so easy to reach the conditions favorable for hopping. In a way, the above-described procedure resembles closely the Monte-Carlo algorithm where the propagation of the system is accepted only when it is advantageous energetically. It is known that Monte-Carlo works well for large sample systems with normal, random distribution of scattering events, thus, by inverting argument we deduce that our system is applicable for a large set of long SWNT shells with normal random distributions of \emph{inter}-shell hopping paths among them. The randomness is in a jumps between shells, while the motion along the MWNT is coherent.

To understand this regime in a more detail, we can look at microscopic origin of the $t_{\perp}$. To this end we introduce electronic wave-function in a given $n-$th shell $\Psi_n(\vec{r})$. We assume that the process of perpendicular hopping spans over a finite time $\tau$ and that during this time the waves in the two involved shells are co-propagating, which allows to define the $t_{\perp}(n,n')$ as the following integral:
\begin{equation}
    t_{\perp}(n,n')=\int_{x-v(x)\tau/2}^{x+v(x)\tau/2}dx d\vec{r}_{\perp} \Psi^*_n(x,\vec{r}_{\perp})\Psi_n'(x,\vec{r}_{\perp})
\end{equation}
The $\vec{r}_{\perp}$ is the coordinate along the circumference of the shell and the dependence $\Psi(\vec{r}_{\perp})$ will be discussed in detail in the later section. Here we focus on the dependence along the nanotubes' axis $\Psi(x)$ which, based on observations of Friedel oscillations in similar systems\cite{odavic2020friedel, das2020friedel}, are expected to be a plane wave. We aim to compute their overlap along the tube. We take two plane waves and integrate over a finite overlap distance $r_{||}$ (and over a possible phase shift between the two waves). The result is presented in Fig.\ref{fig:tperp-calc}.

\begin{figure}
    \centering
    \includegraphics[width=0.69\linewidth]{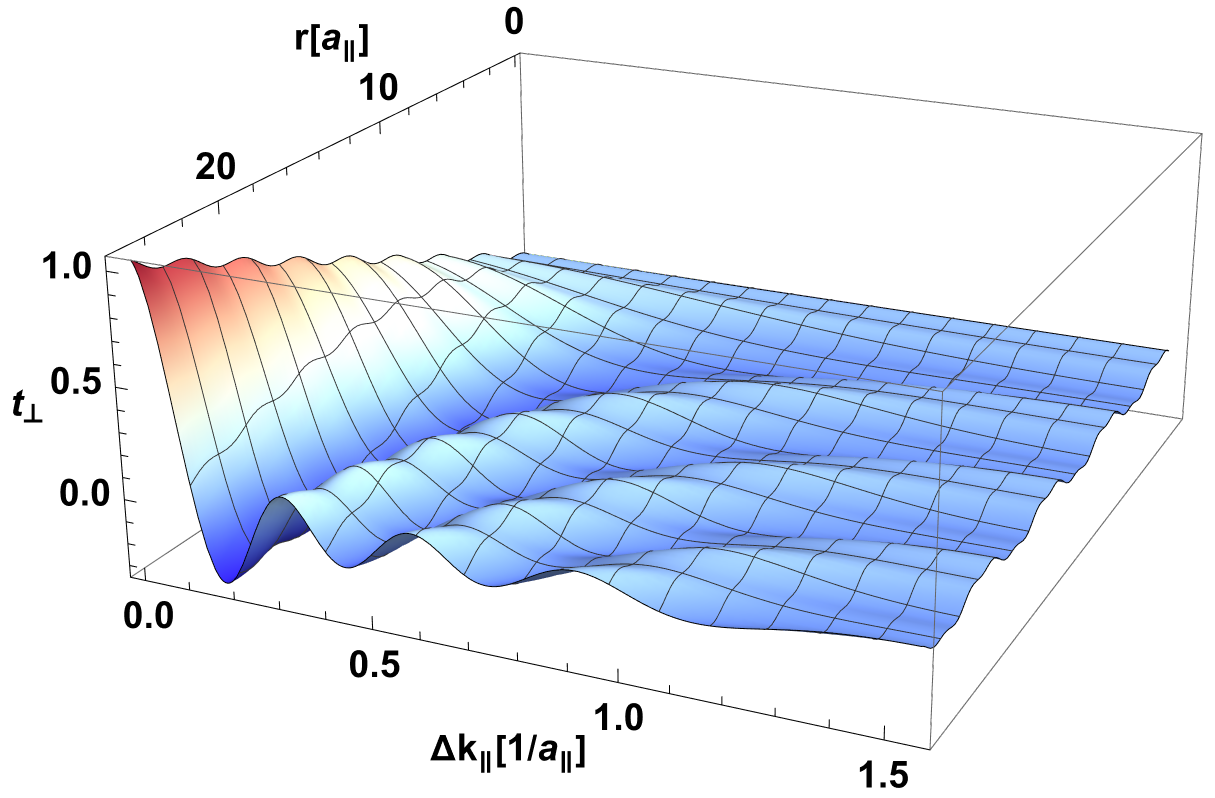}
    \caption{The normalized perpendicular (inter-shell) hybridization $t_{\perp}$ as a function of a momentum misfit $\Delta k_{||}$ of two plane waves and their interaction span $r_{||}$ (components along the tubes axis for both quantities) given in units of lattice constant $a_{||}$ and its inverse, respectively.}
    \label{fig:tperp-calc}
\end{figure}

In the obtained results we observe that as time of co-propagation $\tau=r_{||}/V_F(x)$ increases the peak in the vicinity of $\Delta k_{||}$ increases and narrows. This narrowing implies that momentum is effectively conserved during each hybridization event. For larger $\Delta k_{||}$ we observe decaying oscillatory behavior. Furthermore, anticipating the many-body approach, in the TLL formalism the misfit momenta $\Delta k_{||}$ will be re-distributed among bosonic modes, thus including many body effects will lead to a convolution with Lorentzians with a width $\propto \Delta k_{||}$, thus erasing these higher order peaks.

We now wish to incorporate electron-electron interactions in this description. Since we have weak and extended perturbation, the valley degree of freedom remains a good quantum number also in MWNT. Thus, the TLL description is expected to be just as before, however with unknown values of TLL $K_{\nu}$ parameters. The aim of the paper is thus as follows: based on the past estimates of TLL parameters in SWNT, to derive the value of TLL parameters for MWNT for our model with a random normal distribution of conducting paths among the shells along the length of the tube.  

\section{Values of TLL parameters in SWCNT}\label{sec:KinSWCNT}

There is a rich literature discussing the problem of how to theoretically obtain the values of the TLL parameters in SWCNT. The most comprehensive study was performed in Ref. \cite{egger1998correlated} and we review these results here.

The density-density long-range Coulomb-type interactions between electrons in a SWNT can be expressed as:

\begin{multline}\label{eq:Coul-def}
    H_{Coul}=\sum_{k,q,\mu} [c_{\mu}^{\dag}(k)c_{\mu}(k)]V_{Coul}(k,k')[c_{\mu}^{\dag}(k+q)c_{\mu}(k+q)] \\ 
    = \int dxdx'\nabla\phi_{\rho+}(x)V_{Coul}(x-x')\nabla\phi_{\rho+}(x')
\end{multline}

which we have now expressed both in the fermionic second quantization language and in the bosonic field language. In SWNT, the interaction amplitude $V_{Coul}(k,k')$ does not depend on $\nu$ which is in fact a manifestation of a perfect $C_{\infty}$ cylindrical symmetry of wave functions in SWNT. In the bosonic language, it is clear that only the $\rho+$ mode is affected, which is due to the fact that this mode contains the electric charge.

\begin{figure}
    \centering
    \includegraphics[width=0.39\linewidth]{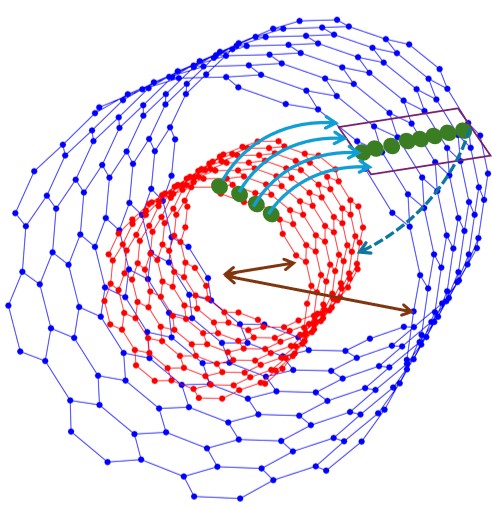}
    \includegraphics[width=0.59\linewidth]{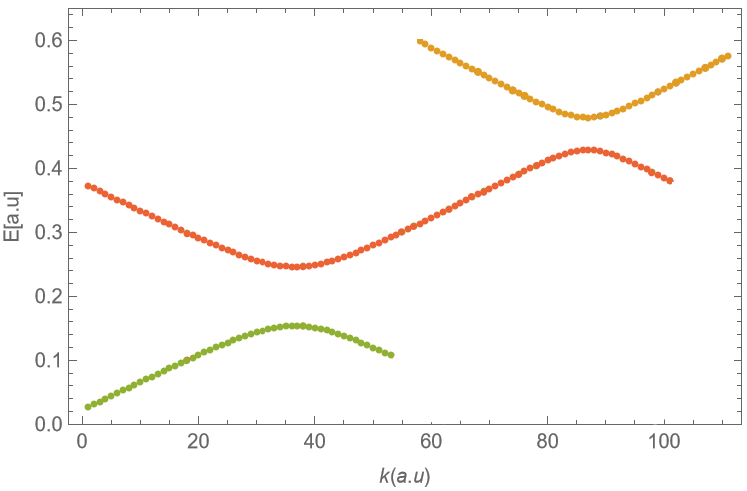}
    \caption{Multi-wall carbon nanotube, illustrating electron hopping process between shells: \textbf{Left,} in real space with inter-shell hopping (blue arrows), and resulting wave-function $\Psi(\vec{r})$ spread only over a finite section (purple parallelogram) on the shell's circumference; dark brown arrows indicate the $R_i$ radius of given shell. \textbf{Right,} in reciprocal space, where we observe a sequence of anti-crossings between bands from different shells. Please, note that both on horizontal and vertical axes we use arbitrary units (in the former case, it is, in fact, a discrete state index)}
    \label{fig:MWNTmod}
\end{figure}

Ref. \cite{egger1998correlated} provides an extensive derivation of Coulomb interactions in a nanotube. It has been found that upon integrating out (evenly) all degrees of freedom along the SWCNT circumference, the following formula for interaction is obtained:

\begin{equation}\label{eq:SWNT-Coul}
V_0(x) = \frac{2e^2}{\kappa \pi \sqrt{a_{z}^{2} + x^2 + 4R^2}} K \left( \frac{2R}{\sqrt{a_{z}^{2} + x^2 + 4R^2}} \right)
\end{equation}

where $K()$ is an elliptic integral of the first kind, $R$ is the radius of given SWNT, and following the Ref. \cite{egger1998correlated}, $a_z  \simeq a_B  \approx a$ (with the Bohr radius $a_B = \sim 2/me^2 = 0.529 \AA$)  denotes the average distance between a $p_z$ electron and the nucleus. In reciprocal space long-wavelength limit its form is:

\begin{equation}\label{eq:CNT-interaction}
    V_{Coul}(q)=\frac{2e^2}{\kappa}(|ln(q R)|+c_0)
\end{equation}

where $\kappa$ is electric permittivity of the nanotube and $c_0=\gamma+\pi/2ln2=0.51$ is a constant, with $\gamma$ -- Euler's constant. This led the authors to the desired formula for $K_{\rho+}$:

\begin{equation}\label{eq:EG-res-K}
    K_{\rho+}=1/\sqrt{1+\frac{8}{\pi}\cdot 2.7\cdot ln(R/L)}
\end{equation}

where the numerical parameter $\frac{2e^2}{v_{\rho+}\kappa}=2.7$ was found by the authors through comparison with experimental data for SWCNT. They suggested value $K_{\rho+}\approx 0.18$ for SWCNT. Clearly, $K_{\rho+}\ll 1$, so it is in a strongly correlated regime and it does depend on the geometrical parameters of the tube. One expects quite substantial variation as a function of these parameters which is indeed observed in SWCNT. 

As for the other bosonic modes, in Ref. \cite{egger1998correlated}, the authors showed that:

\begin{equation}\label{eq:neutral}
    K_{\nu}(x_i)\approx 1 - \sum_{y_i} \nu f(x_i,y_i)
\end{equation}

where $\nu=\pm 1$ for $\nu=\sigma+, \rho-$, respectively. The $x_i$ are some microscopic parameters that determine the geometry of the electron wave function. The $f$ is a function, proportional to a deviation from a perfect circumferential symmetry ($C_{\infty}$ symmetry w.r.t. nanotube axis). Therein, the authors considered only deviations due to atomic positions $(x_i,y_i)$, assuming that otherwise the wave function is perfectly spread over the circumference of the tube. Here, we shall abandon this assumption (see Sec.\ref{ssec:neutral-modes}).

\section{Values of TLL parameters in MWNT}\label{sec:KinMWNT}

\subsection{The holon mode}\label{sec:holon_mode}

However, in MWCNT, the situation is different. As mentioned in Sec. \ref{sec: model}, the MWCNT was introduced as a multi-shell system, with electrons randomly jumping from one SWNT to another within the MWCNT, always choosing the most conducting shell in a given region (Eq. \ref{Eq: Hamiltonian_inter-shell hoppings}). This model is in agreement with the previous experimental \cite{schoenenberger1999interference, stojetz2004ensemble, stojetz2005effect, li2005multichannel} and theoretical \cite{BONARD2001893, bellucci2005crossover, egger2001bulk, abrikosov2005electronic, shuba2009theory} studies. Thus, $R$ is a random variable, and in fact, the propagation length $L$ is random as well, and the variance can be as large as $R$ itself $\delta R/R \propto 1$. One expects a normal, Gaussian distribution of these radii. Therefore, in Eq. \ref{eq:CNT-interaction}, one has to take a logarithm of a Gaussian distribution $\mathcal{G}(R)$ with variance $\delta R$. The solution to this problem is known from the theory of normal distributions. In the $q \rightarrow 0$ limit, the case contributing to $K_{\rho+}$, we have $ln(\mathcal{G}(R))\rightarrow ln(2\pi)=1.84$, which, upon substitution to formula for $K_{\rho+}$, gives:

\begin{equation}\label{eq:Krho-plus-mwnt}
    K_{\rho+}\rightarrow 1/\sqrt{1+\frac{8}{\pi}\cdot2.7\cdot\frac{v_{\rho+}^{swnt}}{v_{\rho+}^{mwnt}}\cdot1.84} = 0.225
\end{equation}

where, like in Eq. \ref{eq:EG-res-K}, we substituted $\frac{2e^2}{v_{\rho+}^{swnt}\kappa}=2.7$. The ratio $\frac{v_{\rho+}^{swnt}}{v_{\rho+}^{mwnt}}=1.48$ can be estimated from the experimentally measured \cite{saqib2018natural} 
ratio of thermal conductivities $\mathfrak{K}_{th}^{swnt}/\mathfrak{K}_{th}^{mwnt}=6600/3000$ if one remembers that $\mathfrak{K}_{th}\propto v_{\rho+}^2$. This is a \emph{quasi-universal} value due to large variations of shell radius, in the sense that it can be modified only if the velocity of carriers $v_{\rho+}^{mwnt}$ changes, for instance at higher temperature due to easier activation of inter-shell hopping the $v_{\rho+}^{mwnt}(T)$ will increase bringing the value even closer to $K_{\rho+}=0.25$. The value $K_{\rho+}=0.25$ is a critical value for quarter-filling 1D Mott insulator (see below). On the other hand, moving towards lower temperatures, one then expects that gradually $v_{\rho+}^{mwnt}(T\rightarrow 0)\rightarrow 0$ thus rapidly decreasing value $K_{\rho+}(T)$. Thus also enhancing the propensity of the system towards any-order charge localization, like pinning, Wigner crystalization which are both potential mechanisms behind the intrinsic Coulomb blockade.

A further correction to $V_{Coul}$ due to variance reads:

\begin{equation}\label{eq:corr1}
    \delta V_{Coul}(q)= - ln(\delta R)
\end{equation}

and is momentum \emph{independent}. The next order correction $\sim q^2$ is expected to be smaller (when $\delta R$ is substantial), and will also be counteracted by Fock exchange corrections proportional to gradients of density (so-called GGA corrections), thus also $\propto q^2$, but with a minus sign.

The momentum-independent component, Eq. \ref{eq:corr1}, will introduce both small and large momentum contributions, the former one modifies the TLL parameter by $\Delta K_{\rho+}$. To make further progress, we note that the large momentum exchange component of electron-electron interactions (which has been neglected so far) leads to non-linear terms $\sim \cos\phi_{\nu}$, the most relevant of which are quarter-filling (two-site unit cell) umklapp terms:

\begin{equation}\label{eq:umkl-CNT}
    H_{umkl}=\int dx g_3(x)\cos2\phi_{\rho+} 
\end{equation}

where $g_{3}$ is the amplitude of umklapp (LL $\rightarrow$ RR) scattering. These terms are not captured by the TLL, but can be incorporated in low energies by gradually averaging higher energies. This is the so-called Renormalization Group procedure, a method the canonical result of which, the Kosterlitz-Thouless flow, is well known \cite{kosterlitz2016kosterlitz, ortiz2013berezinskii, jensen2003kosterlitz, sarkar2021study}. 


Without going into details of the method, which is beyond the scope of this work dedicated more to applications of 1D materials, we note that the $K_{\rho}^*=0.25$ is a special point of this flow -- the system flows towards it in a straight line. To be precise, whenever the deviation of $K_{\rho+}$ from $K_{\rho}^*$, the $\Delta K_{\rho+}$ and the value of $g_3$ are close to each other, then we are in the vicinity of the straight line of the RG flow, thus indeed flowing towards $K_{\rho+}^*=0.25$. It should be emphasized here that in our problem, the high energy is the bandwidth of carbon $p_z$ orbitals, which in SWCNT is above $3eV$, while in MWCNT, it is reduced, for instance, by half (it can be more, each situation can be captured by our theory). At the same time, the energies/temperatures at which MWNT devices work, are around $30meV$ which is smaller by a factor of $50$. Thus, what one measures, is the value of $K_{\rho+}$ \emph{during} the RG flow. From the analysis of logarithmic corrections to correlation functions in TLL, Ref. \cite{giamarchi1989correlation}, 
it is known that the measured value will be indeed $K_{\rho+}=0.25$. 

There are, of course, also Coulomb interactions between the nanotubes. The "other-than-the-metallic" shells of MWCNTs will provide screening, which makes the problem immensely complicated. However, two definite statements can be made:
\begin{itemize}
    \item following the argument given by H. Schulz in Ref. \cite{HJSchulz_1983}, again only the $\phi_{\rho+}$ bosonic mode will be affected, and the entire effect can be captured by minor modification of $K_{\rho+}$,
    \item due to variations of the metal shell positions (that is, how deep inside MWCNT the metallic shell is), the screening is also random, which brings us back to the central limit theorem argument given above.
\end{itemize}

Overall, a very non-trivial result has been obtained here: although the $K_{\rho+}$ parameter is the most affected by interactions and naively has the strongest dependence on geometry, due to generic disorder present in MWCNT, it is possible to simplify the problem of computing it. Contrary to the case of SWCNT, in MWCT in the random hopping regime, the measured value of this parameter turns out to be universal.

\subsection{The three neutral modes}\label{ssec:neutral-modes}

This distinction between SWCNT and MWCNT prevails also in values of parameters of other TLL modes. Any deviation from a perfectly symmetric situation -- even summation over all $\mu$ in Eq. \ref{eq:Coul-def} -- will modify the other three TLL parameters. While SWCNT was generically very symmetric, in MWCNT, the interactions between shells can break the central axis rotational symmetry. The neutral $K_\nu$ modes, $\nu = \sigma_{\pm}, \rho_{-}$, can be modified by external fields, which are defined as fields external to the metallic shell. They can be either generated outside MWCNT in the laboratory or induced due to the presence of other shells. For instance, to modify the spin channel compressibility $K_{\sigma +}$, a field that couples with total spin density -- namely, the local magnetic field -- is required. To modify the relative charge mode TLL parameter (i.e., the compressibility of this mode) $K_{\rho-}$, a force acting differently on two sites of bi-partite lattice is needed, such as a local strain or local dipolar moment. Recent numerical experimental findings \cite{mishra2018coaxial, yuan2017atomistic} show that such forces can indeed be induced in a double-wall CNT. 

The density-density long-range Coulomb-type interactions between electrons in a MWNT need to be expressed in a more general form:

\begin{multline}\label{eq:Coul-def2}
    H_{Coul}=\sum_{k,q,\mu} [c_{\mu}^{\dag}(k)c_{\mu}(k)]V_{Coul}^{\nu\nu'}(k,k')[c_{\mu}^{\dag}(k+q)c_{\mu}(k+q)] \\ 
    = \int dxdx'\Big[\nabla\phi_{\rho+}(x)V_{Coul}^{ \overline{\overline{\nu\nu'}}}(x-x')\nabla\phi_{\rho+}(x')+\sum_{\nu=\rho-,\sigma+}\nabla\phi_{\nu}(x)\delta V_{Coul}(x-x')\nabla\phi_{\nu}(x') \Big]
\end{multline}


where $V_{Coul}^{ \overline{\overline{\nu\nu'}}}(x-x')$ is an interaction averaged over all spin-valley degrees of freedom, while $\delta V_{Coul}(x-x')=\delta V_{Coul}^{\nu\nu'}(x-x')-V_{Coul}^{ \overline{\overline{\nu\nu'}}}(x-x')$ is a deviation from this value. In the following, we shall take: 

$$V_{Coul}^{ \overline{\overline{\nu\nu'}}}(x-x')=V_{C\infty}(x-x')$$

i.e., the perfectly symmetric case gives a good approximation for the average. The expression for the Coulomb interactions in the perfectly symmetric case is the one that had been derived in Ref. \cite{egger1998correlated} for SWCNT, it has been given before, in Eq. \ref{eq:SWNT-Coul}.

In order to predict how the neutral TLL parameters depend on the characteristic features of the material, a deeper understanding of the interactions is required. As already mentioned in the previous section in Ref. \cite{egger1998correlated}, the authors showed that values of the neutral parameters depend on the circumferential symmetry breaking, however now it will be due to the shape of the MWNT wave function $\Psi(\vec{r})$:


\begin{equation}\label{eq:neutral}
    K_{\nu}[\Psi(\vec{r})]\approx 1 - \sum_{\vec{r}}\nu f[\Psi(\varkappa_i)]
\end{equation}

where $\nu=\pm 1$ for $\nu=\sigma+, \rho-$, respectively. The $\varkappa_i$ are some microscopic parameters that effectively determine the geometry of electron wave function $\Psi$. The $f$ is a dimensionless functional proportional to a deviation from a perfect circumferential symmetry, the $C_{\infty}$. It is a ratio:
\begin{equation}\label{eq:f-def}
   f(\varkappa_i)=V_{real}(q\rightarrow 0;\varkappa_i)/V_{C\infty}(q\rightarrow 0)-1 
\end{equation}
We aim to evaluate it now. To this end, we need to carefully examine how the interaction amplitudes are computed when one moves to the second quantization description \cite{altland2010condensed}, the first line in Eq. \ref{eq:Coul-def2}. The scattering amplitude (i.e., a quantity that enters into the second quantization Hamiltonian) of the Hartree-type interaction is given by an integral over the elementary cell: 

\begin{equation}\label{eq:2nd-quant-ampl}
    V_{Coul}^{\mu \mu'} (\vec{r},\vec{r'})= \int d\vec{r} \Psi_{\mu}^{*}(\vec{r})\Psi_{\mu'}^{*}(\vec{r'})\bar{V}_{\rm Coul}(\vec{r}-\vec{r'})\Psi_{\mu}(\vec{r})\Psi_{\mu'}(\vec{r'})
\end{equation}

where the bare Coulomb potential in a nanotube of radius $R$ and thickness $d_z$ reads:

\begin{equation}\label{eq:Coulomb1}
    \bar{V}_{\rm Coul}(\vec{r}-\vec{r'}) = \frac{e^2/\kappa}{\sqrt{(x-x')^2+4R^2\sin^2((y-y')/2R)+d_z^2}}
\end{equation}

where the $d_z$ is the thickness of the toroid (here we shall consider a physical nanotube as a geometrical, stereometric solid figure -- a toroid).

\begin{figure}
    \centering
    \includegraphics[width=0.3\linewidth]{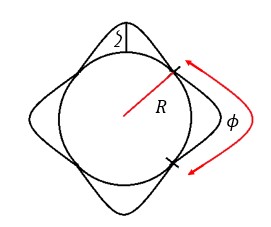}
    \caption{Cross-sectional schematic representation of a carbon nanotube with deformation characterized by parameters $\zeta$ and $\phi$. }
    \label{fig: zeta_phi}
\end{figure}

The $\Psi_{\mu}^{*}(\vec{r}),\Psi_{\mu'}^{*}(\vec{r'})$ are the wave-functions of interacting electrons; they are Bloch waves along the x-direction. In a perpendicular plane, the density, $\rho_{\mu}(\vec{r})=\Psi_{\mu}^{*}(\vec{r})\Psi_{\mu}^{*}(\vec{r})$, can be approximated as a section of a toroid of a varying thickness (the charge density in MWNT is, in general, spread over a section of a distorted toroid). This can be inferred based on Fig. \ref{fig:MWNTmod}, where we see an electron hopping from one shell to another with a charge distribution localized only in a section of the circumference. Furthermore, there will be interference phenomena of waves of different chiral vectors. We then add an extra parameter $\zeta$ that accounts for inhomogeneity along the circumference of the toroid, where $\zeta=0$ corresponds to a symmetric homogeneous distribution (or constant radius), like in a SWCNT. Hence, we shall generalize the expression given by Egger and Gogolin in Ref. \cite{egger1998correlated} for the symmetric SWCNT. We integrate over perpendicular coordinates to get an interaction amplitude along the b-axis $V(x)$:

\begin{equation}\label{eq:toroid}
    V_{real}(x)=\int_{\phi R}^{2\pi R}\int_{\phi R}^{2\pi R} \frac{dy}{2\pi R} \frac{dy'}{2\pi R} \frac{\bar{V}_{\rm Coul}(\vec{r}-\vec{r'})}{1-\zeta\sin((y-y')/2R)}
\end{equation}

where the denominator accounts for different weights introduced by the uneven distributions $\rho_{\mu}(\vec{r})$. The case $\zeta=0$ corresponds to scattering independent on $\mu,\mu'$ (only the uniform component of all circumferential harmonics is present), while $\zeta\neq 0$ corresponds to an appearance of higher circumferential harmonics, as $\zeta\neq 0$ introduces an extra term proportional to $\sin((y-y')/2R)$. In the language of Ref. \cite{egger1998correlated}, the authors proposed the asymmetry due to different atomic positions $x_i$ of two different graphene sub-lattices; here, such difference between sub-lattices appears due to the presence of such higher harmonic component (note: if the distribution is uniform, $\zeta=0$, it remains the same on both sub-lattices). The integral, Eq. \ref{eq:toroid}, can be performed in a closed form also in this more general case:

\begin{equation}\label{eq:inter-eff-real}
    V_{real}(x-x')=\bar{U}\frac{\left(\phi ^2+1\right) \Pi \left(\phi ;\zeta \left|\left(\frac{2 R}{\sqrt{d^2+4 R^2+(x-x')^2}}\right)^2\right.\right)}{\left(\zeta \sqrt{d^2+4 R^2+(x-x')^2}\right)}
\end{equation}

where $U$ is a pre-factor that quantifies the strength of local electron-electron interactions, and $\Pi(\phi;\zeta|1/\tilde{x})$ is the incomplete elliptic integral of the third kind, $\tilde{x}=(x-x')/R$, the relative distance in $R$ units. The integral is parameterized by $U$ (chosen appropriately depending on screening in a given MWNT). The parameters $\phi$ (angle of the sector of the toroid) and $\zeta$ (distortion of the toroid), which are in fact the $\varkappa_i$ in Eq.\ref{eq:f-def}, can be determined by material-specific considerations, namely:
\begin{itemize}
    \item parameter $\zeta$ captures the situation shown in Fig. \ref{fig: zeta_phi}, where the density of electrons along the circumference, related to $\Psi(y,z)$, is not constant. This may be either a static effect induced by Moire-type potential from other shells, or a dynamic effect induced by exciting a phonon (or both). In this second case, there is a possibility to tune the amplitude of $\zeta$ by adjusting the amplitude of IR radiation applied to the emitter. In general, it can be modified by a local stress field; 
    \item parameter $\phi$ is related to the fact that electrons can move only within a part of the nanotube's circumference. One can easily imagine that such a phenomenon will be induced by impurities evaporated on the surface of MWCNT. Then, by adjusting the concentration of impurity atoms, experimentalists and engineers should be able to decrease the average $\phi$, thereby decreasing the sector of the toroid available for mobile electrons $\propto \Psi(y,z)$. Importantly, this effect depends on the chirality of the nanotubes:
    \begin{itemize}
        \item when the metallic nanotube (shell of MWCNT) is armchair (achiral), its wave function $\Psi(y,z)$ forms a simple standing wave with a node on the circumference, meaning that the equation $\Psi(y_0,z_0)=0$ can be fulfilled. The position of the node can be adjusted to the position of the impurity, so for small concentrations of impurities, there will be no effect
        \item when the metallic nanotube (shell) is zig-zag (achiral), it features a uniform wave on the circumference $\Psi(y,z)$ but allows for the degenerate standing wave solution along the nanotube that could admit $\Psi(x\pm a/2)=0$. The atomistic disorder may thus be avoided (by the same argument as above), but the fermionic velocity $v_F$ can be reduced
        \item for the chiral tubes, the wave function $\Psi(x,y,z)$ is a plane wave running in a screw motion along the nanotube. The impurity cannot be in general avoided (although for armchair-like tubes with two $K, K'$ points located at finite $\pm q_0$, there is some adjustment possible, so the influence of atomistic disorder can be weaker). In general, we expect that the smaller the chiral angle, the closer the plane wave will move to the nanotube axis. Thus, the "shadow" of impurity will extend along a longer section of the nanotube.
    \end{itemize}
Of course, as the electron jumps from one least conducting shell to another, then the $\phi$ will vary, but only up to a limited extent because the electron-electron interactions forbid squeezing the wave function $\Psi(\vec{r})$ (and thus the density $\rho(\vec{r})=\Psi^*(\vec{r})\Psi(\vec{r})$) too much. The same applies for the $\zeta$ parameter. 
\end{itemize}

Finally, we note that by definition (Eq. \ref{eq:f-def}), the $f(\zeta,\phi)$ any variation of interaction induced by $\delta \zeta$ or $\delta \phi$ are proportional to $\partial V_{Coul}/V_{Coul}\equiv \partial [Ln[V_{Coul}]]$, thus they are logarithmically suppressed. It thus suffices to take the average values of both parameters in the above formula, Eq.\ref{eq:toroid}. 

\begin{figure}[h]
    \centering
    \includegraphics[width=0.45\textwidth]{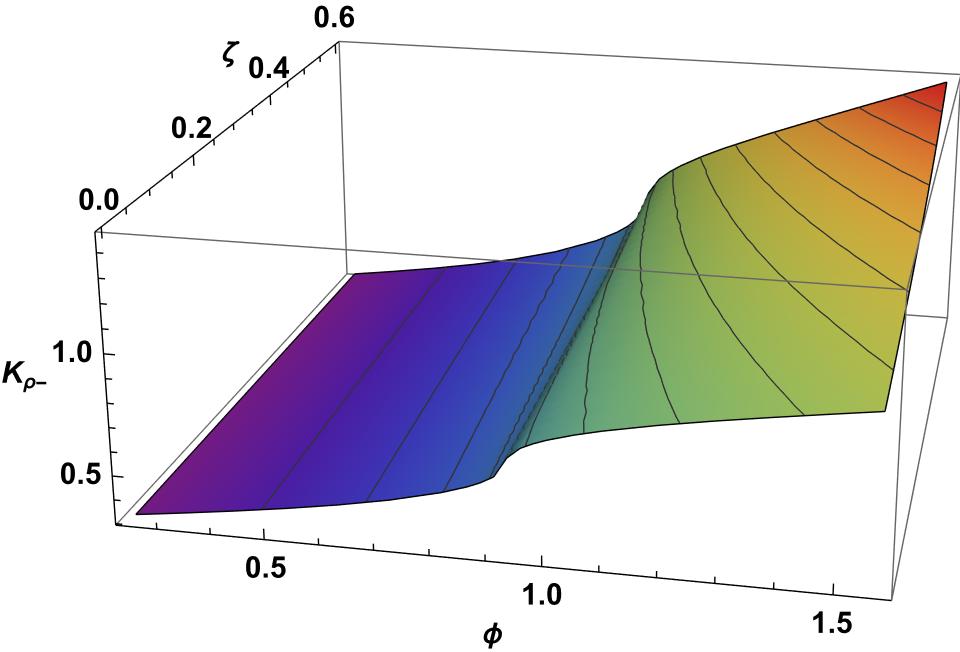} \hspace{1 cm}
    \includegraphics[width=0.45\textwidth]{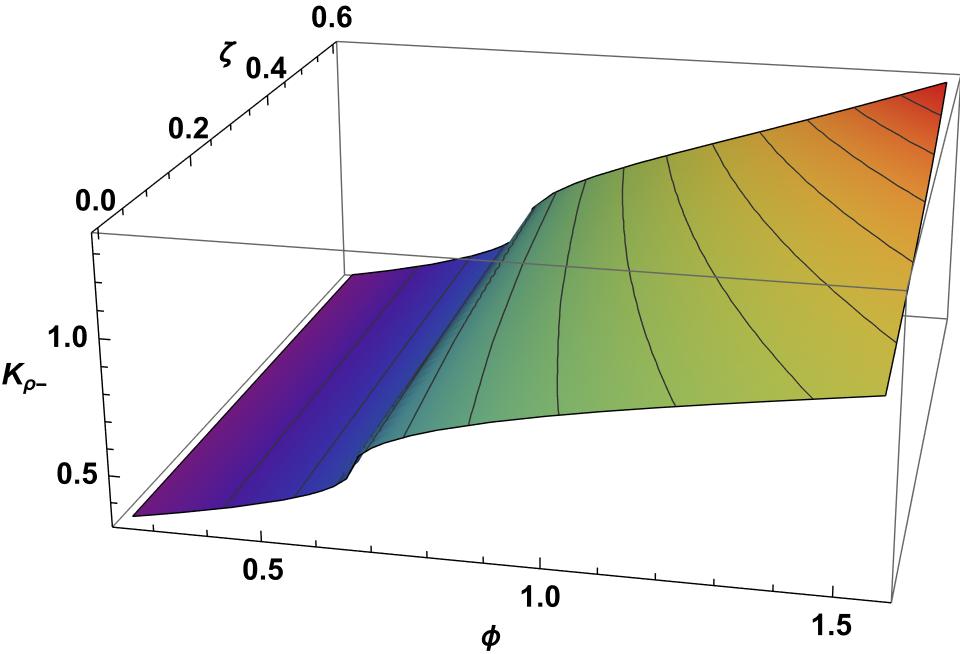}
    (a) \hspace{6.5 cm} (b)\\[0.3 cm]
    \caption{The "neutral", chargeless TLL parameter $K_{\rho-}$ from Eq. \ref{eq:f-explicit}, plotted as a function of the angular span $\phi$ (in radians) of the CNT electronic wave function and its inhomogeneity parameter $\zeta$. In panel (a), we use $a_0=R\left( \left(1-\frac{1}{\sqrt{5}}\right)+1 \right)$ and in (b), $a_0=R\left( 3 \left(1-\frac{1}{\sqrt{5}}\right)+1 \right)$ where $R$ is an average radius of the MWCNT.}
    \label{fig: Chargeless TLL K_rho-_1_3}
\end{figure}

\begin{figure}[t!]
    \centering
    \includegraphics[width=0.48\textwidth]{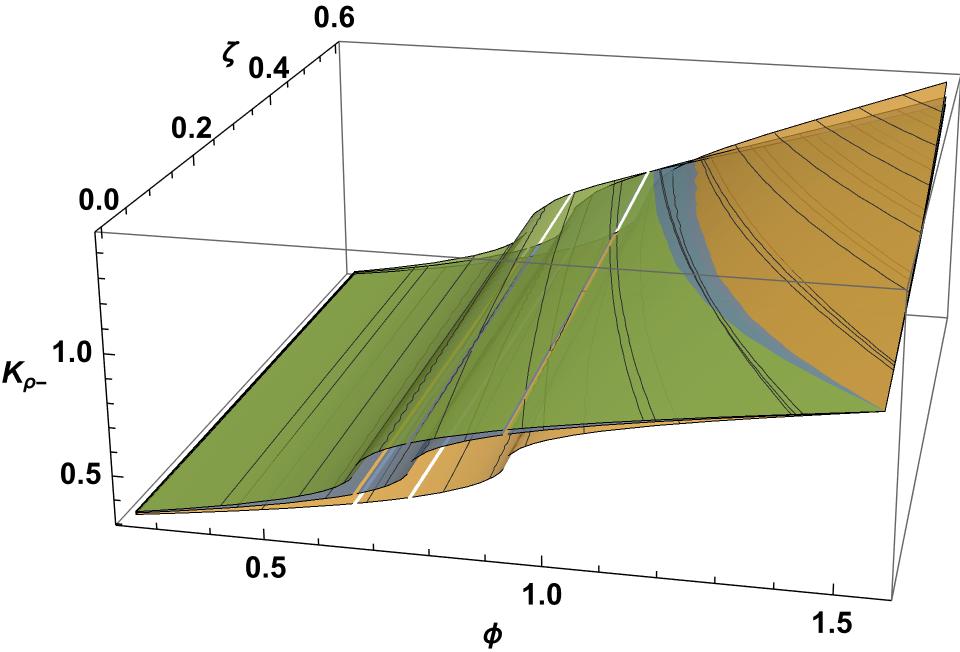}
    \caption{The "neutral", chargeless TLL parameter $K_{\rho-}$ from Eq. \ref{eq:f-explicit}, plotted as a function of the angular span $\phi$ (in radians) of the CNT electronic wave function and its inhomogeneity parameter $\zeta$. We show them here together to visualize the role of $a_0=R\left( j \left(1-\frac{1}{\sqrt{5}}\right)+1 \right)$  variations; orange: $j=1$, blue:  $j=2$, green: $j=3$. }
    \label{fig: Chargeless TLL K_rho-_j}
\end{figure}

The Eq. \ref{eq:CNT-interaction}, which was used to estimate $K_{\rho+}$, is in fact, a long-wavelength approximation for a Fourier transform of the elliptic integral of the first kind $K(x)$, the result of reasoning in Ref. \cite{egger1998correlated} done for the symmetric case $\zeta=0$. Thus, parameter $f$ in Eq. \ref{eq:neutral} is in fact a ratio of the elliptic function of the third and of the first kind, namely:
\begin{equation}\label{eq:f-explicit}
    f(\phi, \zeta) = (1-K_{\rho +})\frac{\Pi(\phi;\zeta|1/\tilde{x})}{K(\tilde{x})} \Bigg|_{\tilde{x} = a_0} - 1
\end{equation}

where $a_0$ is the size of the unit cell along the nanotube (in the conducting shell). In order to fix the value of pre-factor $U$ in Eq. \ref{eq:inter-eff-real}, we multiply the expression by $1-K_{\rho +}$ to ensure that all $K_{\nu}$ parameters deviate proportionally from the non-interacting value $K_{\nu}=1$. The results are presented in Fig. \ref{fig: Chargeless TLL K_rho-_1_3} and Fig. \ref{fig: Chargeless TLL K_rho-_j}.

In Fig. \ref{fig: Chargeless TLL K_rho-_1_3}, we observe that $K_{\rho -}$ is decreasing, thus moving away from the non-interacting value $K=1$,  as $\phi$ angle deviates from $\pi$: in other words, when the wave function spreads only over a finite sector on shell's circumference, the repulsive interactions inside the squeezed wave-packet manifest as decreasing value of $K_{\rho -}$. The influence of $\zeta$ on $K_{\rho-}$ is the opposite: adding a finite value of $\zeta$, thus modulating the charge distribution on the circumference, increases the value of $K_{\rho -}$ pushing it above $K = 1$. The modulation of charge can be thought of as an emergence of a correlation hole, which is now allowed thanks to the lower symmetry of MWNT. Thus, the emergence of the exchange interaction that can indeed raise the value of charge $K_{\rho-}$ (and at the same time favour an in-plane spin configuration thus breaking spin's SU(2) invariance and pushing the value of $K_{\sigma+}$ parameter below one). The last parameter, $a_{0}$,  is essentially the length of the unit cell along the length (the axis of the nanotube), and its role is shown in Fig. \ref{fig: Chargeless TLL K_rho-_j}. This parameter cannot be freely modified, it actually depends on the structure (chirality) of a given shell. We see that this has influence mostly on the intermediate value of the angular sector $\phi$, namely, for shorter lengths of the unit cell, the squeezing effect is then more pronounced. Upon increasing temperature one expects that the electronic wavefunction will become less localized, namely $\zeta(T)$ will decrease and at the same time $\phi(T)$ will increase. By tracing the constant value lines in Fig.\ref{fig: Chargeless TLL K_rho-_1_3} we deduce that will change only a little, with an exception of larger temperatures when $\phi(T)$ will already reach it maximum value, while $\zeta(T)$ will keep decreasing. Then the decreasing value $K_{\rho-}$ will be observed.

\subsection{Measuring $K_{\rho-}$}\label{ssec:measurK}

The main result of this paper is the prediction of the values of neutral parameters $K_{\nu}$ as a function of disorder or strain field. It would be desirable to be able to test this prediction experimentally. Of course, in the 1D regime, the combinations of $K_{\nu}$ will influence all measurable quantities, such as the temperature dependence of transport coefficients. The challenge is to find the quantity that will depend only on $K_{\rho-}$, i.e., will be able to directly manifest the variation of this parameter.

In this regard, an interesting quantity is the frequency of optical phonons measured in Raman experiments. It is known that the frequency of the phonon in the jellium model depends on the electronic susceptibility of the environment $\chi_{\rho}(q\approx 0)$, namely: 
$$
\omega_{ph}(q\rightarrow 0)=  \Omega_0^{ph}/\sqrt{\varepsilon_{\rho}(q\rightarrow 0)} = \Omega_0^{ph}/\sqrt{\chi_{\rho\pm}(q\rightarrow 0)}
$$

where we used the fact that in the random phase approximation (which, following Dzyaloshinskii-Larkin theorem, is exact in TLL), the environment's dielectric function $\varepsilon=1+\chi^{TLL}$ (we drop subscript TLL in the following and assume that close to the singularity $\chi(q\rightarrow 0))\gg 1$) and we took a uniform response because we probe with photons.

Within the TLL formalism, the uniform part of the density is expressed as a gradient of a respective bosonic field $\rho_{\nu}(x,t;q\rightarrow 0)=\nabla\phi_{\nu}(x,t)$. Then the susceptibility, the correlation function of densities reads:
\begin{equation}
    \chi_{\rho\pm}(x,t;q\approx 0)=\langle\nabla\phi_{\rho\pm}(x,t)\nabla\phi_{\rho\pm}(0,0)\rangle=\frac{K_{\rho\pm}}{2 \pi^2}\frac{x^2-y^2}{(x^2+y^2)^2}\Bigg|_{y\equiv \imath v_{\rho\pm} t}
\end{equation}

we thus observe that the pre-factor that enters in front of susceptibility is simply equal to $K_{\nu}$. Using well-known conformal transformation, we can move to finite temperature case, $\beta=1/T$:
\begin{equation}
    \chi_{\rho\pm}(x,t;T,q\approx 0) = \frac{K_{\rho\pm}}{2 \pi^2}\frac{Sinh(\frac{\pi}{\beta}(x+y_p))Sinh (\frac{\pi}{\beta} (x-y_{-p}))}{[Sinh(\frac{\pi}{\beta}(x+y_p)) Sinh(\frac{\pi}{\beta}(x+y_{-p}))]^2}\Bigg|_{y\equiv \imath pv_{\rho\pm} t}
\end{equation}
where $p=\pm1$, to observe that this statement remains valid also in the physically relevant situation of non-zero temperature. The last expression can be Fourier transformed to obtain a finite value $\chi_{\rho\pm}(\omega_{ph},q\approx 0;T)$ in the form of a combination of hyperbolic \emph{Beta} functions. 

Irrespective of details, the following statement can be made: if upon adding an external force $\mathfrak{f}$ (e.g., strain $\propto\zeta$) only the $K_{\nu}$ parameters are changing, the frequency shift reads:
\begin{equation}
    \Delta \omega_{ph}^{i}(q\approx 0) = \frac{\Omega_0^{ph}}{\bar{\chi}^2_{\rho\pm}(\omega_{ph},q\approx 0)}\bar{\chi}'_{\rho\pm}(\omega_{ph},q\approx 0)\Delta K_{\rho\pm}(\mathfrak{f})
\end{equation}
where the $\bar{\chi}$ is the susceptibility without the $K$ prefactor, thus quantity independent of $\mathfrak{f}$. We thus see that by measuring $\Delta \omega_{ph}(\mathfrak{f})$, one can have a direct access to the $\Delta K_{\rho-}(\zeta)$, i.e., the quantity computed in the current work. 

Thus, the measurement of Raman phonon frequency shift offers an opportunity to see how the compressibility $K_{\rho\pm}$ changes as a function of a strain $\mathfrak{f}\equiv \varepsilon$ or by the presence of atomistic impurities. In most cases, particularly for most phonon branches, the response will be that of the total charge mode, the holon $K_{\rho+}$ which does not depend on $\zeta$ in our model. There could be however some further lattice effects that perturb single-particle dispersion, thus these branches can serve as our baseline. However, there is one optical Raman-active mode in which the atoms of opposite sub-lattices (A and B in an underlying bi-partite graphene lattice) will move in the opposite direction. For this special mode, through $\Delta \omega_{ph}^{odd}$, one will measure the desired response of the $K_{\rho-}$. An example of such a Raman experiment with distinct frequencies' shifts of different modes measured as a function of MWCNT perturbation is presented in Ref.\cite{ARAUJO2017348, Bhalearo-Raman-shift}.

A subtle point in the above reasoning is that while the uniform $\bar{\chi}$ does not depend on $K_{\nu}$, it will depend on velocities $v_{\nu}$. Following the discussion below Eq.\ref{eq:Krho-plus-mwnt} we know that these velocities will depend on temperature. Furthermore, all the modifications, renormalizations of velocities are due to single-particle backscattering effects, that affect evenly all modes. To cancel out these T-dependencies of velocities one can apply the above-described baseline procedure of measuring the \emph{relative} shift of different modes: the quantity 

\begin{equation}
    \mathcal{A}(\mathfrak{f},T)=\Delta \omega_{ph}^{odd}/\Delta \omega_{ph}^{even}
\end{equation}
should allow to exclude this effect and focus on $K_{\nu}$ only. Thus the $\mathcal{A}(\mathfrak{f})$ dependence gives a direct experimental access to the $K_{\rho-}(\zeta)$, while the $\mathcal{A}(T)$ dependence will appear mostly due to $K_{\rho+}(T)$ (following discussion below Eq.\ref{eq:Krho-plus-mwnt}). In particular, because $K_{\rho+}(T\rightarrow 0)\rightarrow 0$, the $K_{\rho-}(\zeta)$ will be more pronounced at lower temperatures.

\section{Discussion and Conclusions}\label{Sec: Discussion and Conclusions}

It is worth noting that the postulated dependence of neutral TLL parameter values $K_{\rho-,\sigma+}$ on the external field is stronger when the carriers keep changing the shells during their motion. In a chiral tube, when a carrier stays for a very long distance on one shell, its wave function will unavoidably spread over the entire circumference. Upon changing the shell at a given point, a new cycle of spreading begins but never finishes. 

Previously, the most advanced effective model of MWNT was the one in Ref. \cite{EGGER2002447}, where the authors put forward the idea of intrinsic Coulomb blockade (ICB) and solved the postulated model numerically. Our solution, which starts from more itinerant carriers, does not rely on numerics: instead, it provides analytical expressions that can be easily benchmarked against external perturbations. It should be emphasized that our model is not in contradiction with ICB. It can happen that at the lowest temperatures, the Coulomb charging energy or mini-gaps open in the spectrum. In the higher temperatures, the itinerant picture postulated here is applicable.

The assumption that we used in this study is that there exists only one, albeit fluctuating path of the highest conductivity in MWNT. The multi-shell system does not enter the 2D regime. This can be justified by the fact that the percolation threshold for bcc slabs with $N=3,4$ layers are $p_c=0.35$ and $p_c=0.32$, respectively \cite{horton2017alloy}. On average, one-third of shells are metallic, so by assuming that we have hopping only between the nearest shells, we see that the system is right at the edge of conductivity. At the lowest temperatures, the 2D cross-over is suppressed by the ICB.      

The main outcome of this work is that we have shown what the implications of the proposed model are, where the description of MWNT is given as a series of co-axial SWNT, with the most conducting shell dominating the parallel transport. We derived an effective theory of TLL parameters in such a case. Finally, we showed the behaviour of a measurable quantity, the chemical potential drop, that could serve as a confirmation of our model. From a broader perspective, the advantage of our model is that it allows us to reconcile two experimental facts: i) the TLL properties observed in MWCNT, being in agreement with two leg-ladder modelling of SWNT; ii) the short scattering length which can be interpreted as a short distance between consecutive jumps in between the shells. This last feature also explains the non-zero effects observed for the magnetic field applied along the nanotube axis, which is expected to be exactly zero for a single shell system. Moreover, the presence of a single conducting path explains why the Thouless scale is not observed in transport experiments.

\bibliography{biblio}

\begin{thebibliography}{10}

\bibitem{iijima1991helical}
Sumio Iijima.
\newblock Helical microtubules of graphitic carbon.
\newblock {\em nature}, 354(6348):56--58, 1991.

\bibitem{kukovecz2013multi}
{\'A}kos Kukovecz, G{\'a}bor Kozma, and Zolt{\'a}n K{\'o}nya.
\newblock Multi-walled carbon nanotubes.
\newblock {\em Springer handbook of nanomaterials}, pages 147--188, 2013.

\bibitem{mceuen2000single}
Paul~L McEuen.
\newblock Single-wall carbon nanotubes.
\newblock {\em Physics World}, 13(6):31, 2000.

\bibitem{d2017graphene}
Aditya D~Ghuge, Abhay R~Shirode, and Vilasrao J~Kadam.
\newblock Graphene: a comprehensive review.
\newblock {\em Current drug targets}, 18(6):724--733, 2017.

\bibitem{boumia2014timoshenko}
Lakhdar Boumia, Mohamed Zidour, Abdelnour Benzair, and Abdelouahed Tounsi.
\newblock A timoshenko beam model for vibration analysis of chiral single-walled carbon nanotubes.
\newblock {\em Physica E: Low-dimensional Systems and Nanostructures}, 59:186--191, 2014.

\bibitem{10.1063/1.1419236}
Won~Bong Choi, Jae~Uk Chu, Kwang~Seok Jeong, Eun~Ju Bae, Jo-Won Lee, Ju-Jin Kim, and Jeong-O Lee.
\newblock {Ultrahigh-density nanotransistors by using selectively grown vertical carbon nanotubes}.
\newblock {\em Applied Physics Letters}, 79(22):3696--3698, 11 2001.

\bibitem{liu2015curvature}
Hong Liu, Dirk Heinze, Huynh~Thanh Duc, Stefan Schumacher, and Torsten Meier.
\newblock Curvature effects in the band structure of carbon nanotubes including spin--orbit coupling.
\newblock {\em Journal of Physics: Condensed Matter}, 27(44):445501, 2015.

\bibitem{dumitricua2002curvature}
Traian Dumitric{\u{a}}, Chad~M Landis, and Boris~I Yakobson.
\newblock Curvature-induced polarization in carbon nanoshells.
\newblock {\em Chemical Physics Letters}, 360(1-2):182--188, 2002.

\bibitem{chudz-so-cnt}
Piotr Chudzinski.
\newblock Spin-orbit coupling and proximity effects in metallic carbon nanotubes.
\newblock {\em Physical Review B}, 92:115147, Sep 2015.

\bibitem{pichugin2014spin}
Konstantin~Nikolaevich Pichugin, Mihal Pudlak, and Rashid~Giyasovich Nazmitdinov.
\newblock Spin-orbit effects in carbon nanotubes--{A}nalytical results.
\newblock {\em The European Physical Journal B}, 87:1--10, 2014.

\bibitem{hamada1992new}
Noriaki Hamada, Shin-ichi Sawada, and Atsushi Oshiyama.
\newblock New one-dimensional conductors: {G}raphitic microtubules.
\newblock {\em Physical review letters}, 68(10):1579, 1992.

\bibitem{dresselhaus2005semiconducting}
MS~Dresselhaus, R~Saito, and A~Jorio.
\newblock Semiconducting carbon nanotubes.
\newblock In {\em AIP Conference Proceedings}, volume 772, pages 25--31. American Institute of Physics, 2005.

\bibitem{jabeen2015review}
Saira Jabeen, Ayesha Kausar, Bakhtiar Muhammad, Sagheer Gul, and Muhammad Farooq.
\newblock A review on polymeric nanocomposites of nanodiamond, carbon nanotube, and nanobifiller: structure, preparation and properties.
\newblock {\em Polymer-Plastics Technology and Engineering}, 54(13):1379--1409, 2015.

\bibitem{giamarchi2003quantum}
Thierry Giamarchi.
\newblock {\em {Quantum {P}hysics in {O}ne {D}imension}}.
\newblock Oxford University Press, 2003.

\bibitem{haldane1981luttinger}
FDM Haldane.
\newblock '{L}uttinger liquid theory' of one-dimensional quantum fluids. i. properties of the {L}uttinger model and their extension to the general 1{D} interacting spinless {F}ermi gas.
\newblock {\em Journal of Physics C: Solid State Physics}, 14(19):2585, 1981.

\bibitem{he2020emergence}
Feng He, Yu-Zhu Jiang, Hai-Qing Lin, Randall~G Hulet, Han Pu, and Xi-Wen Guan.
\newblock Emergence and disruption of spin-charge separation in one-dimensional repulsive fermions.
\newblock {\em Physical review letters}, 125(19):190401, 2020.

\bibitem{PhysRevB.108.L201101}
A.~V. Parafilo, V.~M. Kovalev, and I.~G. Savenko.
\newblock Probing {L}uttinger liquid properties in a multichannel two-site charge {K}ondo simulator.
\newblock {\em Phys. Rev. B}, 108:L201101, Nov 2023.

\bibitem{doi:10.1126/science.abn1719}
Ruwan Senaratne, Danyel Cavazos-Cavazos, Sheng Wang, Feng He, Ya-Ting Chang, Aashish Kafle, Han Pu, Xi-Wen Guan, and Randall~G. Hulet.
\newblock Spin-charge separation in a one-dimensional {F}ermi gas with tunable interactions.
\newblock {\em Science}, 376(6599):1305--1308, 2022.

\bibitem{shen2023effect}
L~Shen, A~Alshemi, E~Campillo, E~Blackburn, Paul Steffens, Martin Boehm, D~Prabhakaran, and AT~Boothroyd.
\newblock Effect of magnetic order on longitudinal {T}omonaga-{L}uttinger liquid spin dynamics in weakly coupled spin-$\frac{1}{2}$ chains.
\newblock {\em Physical Review B}, 107(13):134425, 2023.

\bibitem{horvatic2020direct}
Mladen Horvati{\'c}, Martin Klanj{\v{s}}ek, and Edmond Orignac.
\newblock Direct determination of the {T}omonaga-{L}uttinger parameter {K} in quasi-one-dimensional spin systems.
\newblock {\em Physical Review B}, 101(22):220406, 2020.

\bibitem{10.1093/pnasnexus/pgae363}
Sharath~Kumar Channarayappa, Sankalp Kumar, N~S Vidhyadhiraja, Sumiran Pujari, M~P Saravanan, Amal Sebastian, Eun~Sang Choi, Shalinee Chikara, Dolly Nambi, Athira Suresh, Siddhartha Lal, and D~Jaiswal-Nagar.
\newblock Tomonaga–{L}uttinger liquid and quantum criticality in spin-$\frac{1}{2}$ antiferromagnetic {H}eisenberg chain ${C}_{14}{H}_{18}{C}u{N}_{4}{O}_{10}$ via {W}ilson ratio.
\newblock {\em PNAS Nexus}, 3(9):pgae363, 08 2024.

\bibitem{lange2024metal}
Florian Lange and Holger Fehske.
\newblock Metal-insulator transition of spinless fermions coupled to dispersive optical bosons.
\newblock {\em Scientific Reports}, 14(1):18050, 2024.

\bibitem{cavazos2023thermal}
Danyel Cavazos-Cavazos, Ruwan Senaratne, Aashish Kafle, and Randall~G Hulet.
\newblock Thermal disruption of a {L}uttinger liquid.
\newblock {\em Nature Communications}, 14(1):3154, 2023.

\bibitem{EGGER2002447}
R~Egger and A.O Gogolin.
\newblock Intrinsic {C}oulomb blockade in multi-wall carbon nanotubes.
\newblock {\em Chemical Physics}, 281(2):447--454, 2002.

\bibitem{inaba2022electron}
Takumi Inaba, Takahiro Morimoto, Satoshi Yamazaki, and Toshiya Okazaki.
\newblock Electron scattering by friedel oscillations in carbon nanotubes.
\newblock {\em Nano Research}, 15(2):889--897, 2022.

\bibitem{dini2020unified}
Yoann Dini, J{\'e}r{\^o}me Faure-Vincent, and Jean Dijon.
\newblock A unified electrical model based on experimental data to describe electrical transport in carbon nanotube-based materials.
\newblock {\em Nano Research}, 13:1764--1779, 2020.

\bibitem{su2007work}
WS~Su, Tsan-Chuen Leung, and Che~Ting Chan.
\newblock Work function of single-walled and multiwalled carbon nanotubes: First-principles study.
\newblock {\em Physical Review B—Condensed Matter and Materials Physics}, 76(23):235413, 2007.

\bibitem{egger2001luttinger}
R~Egger, A~Bachtold, MS~Fuhrer, M~Bockrath, DH~Cobden, and PL~McEuen.
\newblock Luttinger liquid behavior in metallic carbon nanotubes.
\newblock In {\em Interacting Electrons in Nanostructures}, pages 125--146. Springer, 2001.

\bibitem{Nanot2013}
Sebastien Nanot, Nicholas~A. Thompson, Ji-Hee Kim, Xuan Wang, William~D. Rice, Erik~H. H{\'a}roz, Yogeeswaran Ganesan, Cary~L. Pint, and Junichiro Kono.
\newblock {\em Single-Walled Carbon Nanotubes}, pages 105--146.
\newblock Springer Berlin Heidelberg, Berlin, Heidelberg, 2013.

\bibitem{mishra2018coaxial}
Brijesh~Kumar Mishra and Balakrishnan Ashok.
\newblock Coaxial carbon nanotubes: from springs to ratchet wheels and nanobearings.
\newblock {\em Materials Research Express}, 5(7):075023, 2018.

\bibitem{ye2023study}
Xuan Ye, Mengxiong Liu, Xide Li, and Xiaoming Liu.
\newblock Study of interwall interaction during the pull separation of ultra-long double-walled carbon nanotubes under lateral loading.
\newblock {\em Extreme Mechanics Letters}, 64:102079, 2023.

\bibitem{CARON1988A67}
L.G Caron and C~Bourbonnais.
\newblock Renormalization of transverse hopping integral in quasi-1{D} conductors.
\newblock {\em Synthetic Metals}, 27(1):A67--A74, 1988.
\newblock Proceedings of the International Conference on Science and Technology of Synthetic Metals.

\bibitem{yu2022interlayer}
Guodong Yu, Yunhua Wang, Mikhail~I Katsnelson, Hai-Qing Lin, and Shengjun Yuan.
\newblock Interlayer hybridization in graphene quasicrystal and other bilayer graphene systems.
\newblock {\em Physical Review B}, 105(12):125403, 2022.

\bibitem{saqib2018natural}
Muhammad Saqib, Ilyas Khan, and Sheridan Shafie.
\newblock Natural convection channel flow of {CMC}-based {CNT}s nanofluid.
\newblock {\em The European Physical Journal Plus}, 133(12):549, 2018.

\bibitem{Carr-dimSWCNT}
Sam~T. Carr, Alexander~O. Gogolin, and Alexander~A. Nersesyan.
\newblock Interaction induced dimerization in zigzag single wall carbon nanotubes.
\newblock {\em Phys. Rev. B}, 76:245121, Dec 2007.

\bibitem{odavic2020friedel}
Jovan Odavi{\'c}, Nicole Helbig, and Volker Meden.
\newblock Friedel oscillations of one-dimensional correlated fermions from perturbation theory and density functional theory.
\newblock {\em The European Physical Journal B}, 93:1--11, 2020.

\bibitem{das2020friedel}
Joy~Prakash Das, Chandramouli Chowdhury, and Girish~S Setlur.
\newblock Friedel oscillations and dynamical density of states of an inhomogeneous {L}uttinger liquid.
\newblock {\em Physica Scripta}, 95(7):075710, 2020.

\bibitem{egger1998correlated}
Reinhold Egger and Alexander~O Gogolin.
\newblock Correlated transport and non-{F}ermi-liquid behavior in single-wall carbon nanotubes.
\newblock {\em The European Physical Journal B-Condensed Matter and Complex Systems}, 3(3):281--300, 1998.

\bibitem{schoenenberger1999interference}
Ch~Schoenenberger, A~Bachtold, Strunk, C, J-P Salvetat, and L~Forro.
\newblock Interference and {I}nteraction in multi-wall carbon nanotubes.
\newblock {\em Applied Physics A}, 69:283--295, 1999.

\bibitem{stojetz2004ensemble}
B~Stojetz, Ch~Hagen, Ch~Hendlmeier, E~Ljubovi{\'c}, L~Forro, and Ch~Strunk.
\newblock Ensemble averaging of conductance fluctuations in multiwall carbon nanotubes.
\newblock {\em New Journal of Physics}, 6(1):27, 2004.

\bibitem{stojetz2005effect}
Bernhard Stojetz, Csilla Miko, Laszlo Forr{\'o}, and Christoph Strunk.
\newblock Effect of band structure on quantum interference in multiwall carbon nanotubes.
\newblock {\em Physical review letters}, 94(18):186802, 2005.

\bibitem{li2005multichannel}
H.~J. Li, W.~G. Lu, J.~J. Li, X.~D. Bai, and C.~Z. Gu.
\newblock Multichannel {B}allistic transport in multiwall carbon nanotubes.
\newblock {\em Phys. Rev. Lett.}, 95:086601, Aug 2005.

\bibitem{BONARD2001893}
Jean-Marc Bonard, Hannes Kind, Thomas Stöckli, and Lars-Ola Nilsson.
\newblock Field emission from carbon nanotubes: the first five years.
\newblock {\em Solid-State Electronics}, 45(6):893--914, 2001.

\bibitem{bellucci2005crossover}
S~Bellucci, J~Gonzalez, and Pasquale Onorato.
\newblock Crossover from the {L}uttinger-liquid to {C}oulomb-blockade regime in carbon nanotubes.
\newblock {\em Physical review letters}, 95(18):186403, 2005.

\bibitem{egger2001bulk}
R~Egger and AO~Gogolin.
\newblock Bulk and boundary zero-bias anomaly in multiwall carbon nanotubes.
\newblock {\em Physical Review Letters}, 87(6):066401, 2001.

\bibitem{abrikosov2005electronic}
AA~Abrikosov~Jr, DV~Livanov, and AA~Varlamov.
\newblock Electronic spectrum and tunneling properties of multiwall carbon nanotubes.
\newblock {\em Physical Review B—Condensed Matter and Materials Physics}, 71(16):165423, 2005.

\bibitem{shuba2009theory}
MV~Shuba, G~Ya Slepyan, SA~Maksimenko, C~Thomsen, and A~Lakhtakia.
\newblock Theory of multiwall carbon nanotubes as waveguides and antennas in the infrared and the visible regimes.
\newblock {\em Physical Review B—Condensed Matter and Materials Physics}, 79(15):155403, 2009.

\bibitem{kosterlitz2016kosterlitz}
J~Michael Kosterlitz.
\newblock Kosterlitz--{T}houless physics: a review of key issues.
\newblock {\em Reports on Progress in Physics}, 79(2):026001, 2016.

\bibitem{ortiz2013berezinskii}
G~Ortiz, E~Cobanera, and Z~Nussinov.
\newblock Berezinskii--{K}osterlitz--{T}houless transition through the eyes of duality.
\newblock In {\em 40 Years of Berezinskii--Kosterlitz--Thouless Theory}, pages 93--134. World Scientific, 2013.

\bibitem{jensen2003kosterlitz}
Henrik~Jeldtoft Jensen.
\newblock The {K}osterlitz-{T}houless transition.
\newblock {\em Lecture Notes on Kosterlitz-Thouless Transition in the XY Model}, 2003.

\bibitem{sarkar2021study}
Sujit Sarkar.
\newblock A study of quantum {B}erezinskii--{K}osterlitz--{T}houless transition for parity-time symmetric quantum criticality.
\newblock {\em Scientific Reports}, 11(1):5510, 2021.

\bibitem{giamarchi1989correlation}
T~Giamarchi and HJ~Schulz.
\newblock Correlation functions of one-dimensional quantum systems.
\newblock {\em Physical Review B}, 39(7):4620, 1989.

\bibitem{HJSchulz_1983}
H~J Schulz.
\newblock Long-range {C}oulomb interactions in quasi-one-dimensional conductors.
\newblock {\em Journal of Physics C: Solid State Physics}, 16(35):6769, dec 1983.

\bibitem{yuan2017atomistic}
Xuebo Yuan and Youshan Wang.
\newblock Atomistic simulations on interwall sliding behaviour of double-walled carbon nanotube: effects of structural defects.
\newblock {\em Molecular Simulation}, 43(12):953--961, 2017.

\bibitem{altland2010condensed}
Alexander Altland and Ben~D Simons.
\newblock {\em Condensed matter field theory}.
\newblock Cambridge university press, 2010.

\bibitem{ARAUJO2017348}
P.T. Araujo, N.M. {Barbosa Neto}, M.E.S. Sousa, R.S. Angélica, S.~Simões, M.F.G. Vieira, M.S. Dresselhaus, and M.A. {Leite dos Reis}.
\newblock Multiwall carbon nanotubes filled with {A}l4{C}3: Spectroscopic signatures for electron-phonon coupling due to doping process.
\newblock {\em Carbon}, 124:348--356, 2017.

\bibitem{Bhalearo-Raman-shift}
G.~M. Bhalerao, M.~K. Singh, A.~K. Sinha, and Haranath Ghosh.
\newblock Optical redshift in the {R}aman scattering spectra of {F}e-doped multiwalled carbon nanotubes: Experiment and theory.
\newblock {\em Phys. Rev. B}, 86:125419, Sep 2012.

\bibitem{horton2017alloy}
MK~Horton and MA~Moram.
\newblock Alloy composition fluctuations and percolation in semiconductor alloy quantum wells.
\newblock {\em Applied Physics Letters}, 110(16), 2017.

\end{thebibliography}

\end{document}